**Evolution of topological defects at two sequential phase transitions of $Nd_2SrFe_2O_7$**

Fei-Ting Huang[1,‡], Yanbin Li[1,2,‡], Fei Xue[3,‡], Jae-Wook Kim[1], Lunyong Zhang[4], Ming-Wen Chu[5], Long-Qing Chen[3], and Sang-Wook Cheong[1*]

[1]Rutgers Center for Emergent Materials and Department of Physics and Astronomy, Rutgers University, Piscataway, New Jersey 08854, USA,

[2]State Key Laboratory of Crystal Materials, Shandong University, Jinan, China,

[3]Department of Materials Science and Engineering, The Pennsylvania State University, University Park, Pennsylvania 16802, USA.

[4]Laboratory for Pohang Emergent Materials and Max Plank POSTECH Center for Complex Phase Materials, Pohang University of Science and Technology, Pohang 790-784, Korea.

[5] Center for Condensed Matter Sciences and Center of Atomic Initiative for New Materials, National Taiwan University, Taipei 106, Taiwan

[*] Corresponding author: sangc@physics.rutgers.edu

[‡]These authors contributed equally to this work.

**How topological defects, unavoidable at symmetry-breaking phase transitions in a wide range of systems, evolve through consecutive phase transitions with different broken symmetries remains unexplored. $Nd_2SrFe_2O_7$, a bilayer ferrite, exhibits two intriguing structural phase transitions and dense networks of the so-called type-II $Z_8$ structural vortices at room temperature, so it is an ideal system to explore the topological defect evolution. From our extensive experimental investigation, we demonstrate that the cooling rate at the second-order transition ($1290^oC$) plays a decisive role in determining the vortex density at room temperature, following the universal Kibble-Zurek mechanism. In addition, we discovered a transformation between topologically-distinct vortices ($Z_8$ to $Z_4$ vortices) at the first-order transition ($550^oC$), which conserves the number of vortex cores. Remarkably, the $Z_4$ vortices consist of two phases with an identical symmetry but two distinct magnitudes of an order parameter. Furthermore, when lattice distortion is enhanced by chemical doping, a new type of topological defects emerges: loop domain walls with orthorhombic distortions in the tetragonal background, resulting in unique pseudo-orthorhombic twins. Our findings open**

**a new avenue to explore the evolution of topological defects through multiple phase transitions.**

# I. INTRODUCTION

Recent progress in observational cosmology is remarkable and testing the limits of our knowledge [1]. In parallel, the ground-based quantized vortices of superfluid [2,3], optical vortices of liquid crystals [4,5], multiferroic vortices [6-9], polarization screws of oxide superlattices [10, 11] and magnetic skyrmions [12] in condensed matter have made compelling analogies to putative cosmological topological defects such as cosmic strings and cosmic domain walls (DWs) [13-14]. Understanding topological defects at phase transition is particularly important in light of their crucial role as fingerprints of order parameter topologies [7, 15] and phase fluctuations at the critical temperature through the Kibble-Zurek mechanism [13,14]. The Kibble-Zurek mechanism describes the scaling of the defect density and how information spreads at a finite speed in a system that is driven through a continuous phase transition. It has been effectively tested on a variety of condensed matter systems such as liquid crystals [16,17], superfluid $^3$He [2] and multiferroics [7,18]. However, little is known about how topological defects evolve upon crossing multiple phase transitions.

The prototypical model used to describe spontaneous symmetry-breaking is associated with a Mexican hat-type potential-energy landscape [19-21], where the distance from the peak and the angle around the axis represent the magnitude and phase of an order parameter, respectively. A system characterized by this model energy landscape may have a continuous $U(1)$ symmetry peak near the critical temperature ($T_c$), which plays a key role in the initial topological vortex-like defect formation at $T_c$ before the system falls to the brim with discrete global minima at low temperatures, leading to discrete vortices. Multiferroic hexagonal manganite is an example of a phase transition described by such an energy landscape [19, 21]. Interestingly, the restoration of a continuous $U(1)$ symmetry has been predicted and confirmed at vortex cores even at low temperatures [20, 22-24], leading to topological protection from low-energy perturbations [25]. Thus, vortex domains are protected in a symmetry sense. This unique topological protection may enable a transformation between two different types of topological vortices. This possibility arises when the regions with the metastable-state symmetry being trapped within the ground-state vortex grow as the metastable-state becomes favored via thermal activation, chemical doping or mechanical means. The left path in Fig. 1b depicts this scenario while the right path illustrates the commonly observed O' orthorhombic twins with 90°-type straight and 180°-type curved DWs. Dark and light colors correspond to the ground state and a metastable state, respectively. In the left path, the brown vortex core acts as a nucleation center, and DWs are associated with local minima, separating the global minima in an energy landscape as shown in Fig. 1c. The vortex structures are preserved when the system switches between global minima and local minima. Thus, for example, metastable-state vortices may evolve from ground-state vortices far below the critical temperature. The first requirement to realize such vortex transformation is to find a system having DWs with metastability that can be controlled.

From our study on $(Nd,Tb,Sr)_3Fe_2O_7$, $n = 2$ Ruddlesden-Popper (RP) ferrite, we have discovered a new vortex-to-vortex transformation, which is unexpected, but consistent with topological protection. Ferrite is an intriguing system that exhibits one high-temperature second-order structural transition and a following first-order structural transition above room temperature in addition to two magnetically ordered states at low temperatures [26-29], so it is an ideal system to explore topological phase transitions systematically within one system. Figures 1a-1b illustrate the details of various relevant structures. First of all, we have discovered the so-called type-II $Z_8$ structural vortices structural vortices at room-temperature and examined the dynamics of $Z_8$ T' structural vortices utilizing in-situ heating dark-field transmission electron microscopy (DF-TEM). We have explored the effects of thermally-induced kinetic processes, long-range spontaneous distortion, and chemical doping on the resulting topological defect patterns and attempted to explain the origin of domain configurations by phase-field simulations. Our results show that, despite multiple phase transitions, the Kibble-Zurek mechanism is validated if the length scale is set at a second-order critical temperature. We have also unveiled an unexpected vortex-to-loop transformation due to the enlarged elastic interaction with chemical doping and thermodynamic and strain/pressure controls of topological defect patterns.

## II. POSSIBLE SCENARIOS OF VORTEX TRANSFORMATION

Intriguing structural type-II $Z_8$ T' structural vortices (eight-state vortex-antivortex pairs) were recently found in $Ca_2SrTi_2O_7$, $n = 2$ RP bi-layered perovskite [30]. They exhibit a unique real-space topology in which domains and DWs are intricately intertwined with two sets of octahedral tilts marked by red and black arrows (the bottom cartoon of Fig. 1b). Figure 2a illustrates those sets of arrows, rotating every 45 degrees, are associated with two distinct symmetries: an orthorhombic O' phase (*Amam*) and tetragonal T' phase (*P*4$_2$/*mnm*), respectively. The smooth 90°-rotating, for example, from an up-tilt to a right-tilt state (Fig. 1b yellow to red domains), occurs via a diagonal-tilt state (light-red DW in Fig. 1b), resulting in a sequence of all consecutive eight states around an un-tilted core in brown (Fig. 2a). Note that four states of the O' phase appear narrow as DWs and the other four states of the T' phase remain broad as domains. This type of domain configurations with large disparity in size is called type-II domains; otherwise, type-I. Emphasize that the symmetry in the order parameter space is 4-fold, but 4 distinct domain states and 4 distinct domain wall states are present around a vortex core, and an oxygen displacement vector relevant to the order parameter rotates 8 times around vortex core, so we call these vortices as type-II $Z_8$ vortices. Experimentally, however, Ca$_2$SrTi$_2$O$_7$ reveal a direct tetragonal T to T' structural transition, indicating a significant instability of the desired orthorhombic O' phase [30].

On the other hand, Nd$_2$SrFe$_2$O$_7$ ferrite, the structural analogy of titanates, was reported to show two consecutive structural transitions with the target orthorhombic O' phase as an intermediate symmetry appearing above 550℃ [26, 32]. It first undergoes a second-order structural transition from the high-temperature T phase to the intermediate orthorhombic O' phase, denoted as H-T/O' transition, at the critical temperature $T_{c1}$ of ~1290℃. A following first-order structural transition, L-O'/T' transition denoted as $T_{c2}$, occurs from the O' phase to the ground-state T' phase. Note that O' symmetry is a subgroup of T symmetry, which is consistent with the second-order nature across $T_{c1}$ while T' symmetry is not a subgroup of O' phase, which is in accordance with the first-order nature of $T_{c2}$. Microscopically, O' and T' phase are described by different tilts of the FeO$_6$ octahedra (Fig. 1a): the T' phase has the tilt axis along the Fe-O bond direction, defined as the [100]$_T$, while the tilt occurs along the diagonal direction, i.e. the [110]$_T$ in the O' phase. Figure 1a illustrates the in-plane structural models in which the black arrowheads show the details of apical oxygen tilts. The tilts of FeO$_6$ octahedra result from the condensation of the $X_3^-$ soft mode. Based on Landau theory, we plotted an energy landscape Fig. 1c as a function of order parameters with polar coordinates ($Q$, $\varphi$). Here, $Q$ and $\varphi$ are the magnitude and phase of the $X_3^-$ mode. With the in-plane axis of the T phase as the polar axis, the phase $\varphi$ can take the value $\frac{n\pi}{4}$ ($n = 1 - 8$).

When *n* is an even number, the system is in the T' phase, while an odd *n* indicates an O' phase. The T' phase and O' phase correspond to the global minima and local minima in the energy landscape, respectively. The energy barriers between the global minima (T') and the local minima (O') is ~65 meV, which is about two times of that in hexagonal $YMnO_3$ [19]. Full computational details [33-37] are given in Supplemental Material section I [31]. In addition to two sequential structural transitions, the system is in-plane antiferromagnetic below ~550 K [27-29] and spins are aligned along the c axis below 15 K (the details are discussed in Supplemental Material section II and Fig. S1 [31]).

## III. THE KIBBLE ZUREK MECHANISM

Figures 3a-3c show our superlattice DF-TEM images in a $Nd_2SrFe_2O_7$ single crystal taken under different tilting conditions and its corresponding diffraction pattern along the $[001]_T$ direction (Fig. 3d). The curved dark-contrast lines reveal boundaries of four bright T'-phase domains merging at one core, which is a non-T' state. Intriguingly, those curved DWs exhibit two distinct extinction rules in the superlattice DF-TEM images (Figs. 3b-3c), in which the inequivalent nature of these two-types DWs indicates an orthorhombic-like local structure at those walls, consistent with Ref [30]. The directions of apical oxygen distortions are denoted as black arrows in the insets of Figs. 3b and 3c associated with $[110]_T$ and $[\bar{1}10]_T$ tilt axes, respectively. First, the domain pattern can be viewed as a cut-through graph of these two types of O'-symmetry DWs (light-red and light-yellow bold lines in Fig. 3e). Note that the relative spontaneous distortion $\pm a_{O'}$ directions can be identified from the related electron diffraction patterns (Fig. 3d), but the absolute distortion direction cannot be. Thus, once the distortion direction is chosen for DWs, then the distortion directions in neighboring domains can be fully assigned without ambiguity. Therefore, oxygen octahedra inside the bright contrast domains tilt along either $[100]_T$ or $[010]_T$ directions whereas those in the dark-contrast DWs tilt along diagonal $<110>_T$ directions. Figure 3f shows one possible mapping of eight variants in the domains and DWs. The vortices and antivortices are categorized based on the cycling sequence of the eight states around the cores. Figure 2 identifies four types of (anti-)vortex domains observed in Fig. 3f.

We found that the vortex density D (core number/area) remains intact with various thermal treatments below 1280°C, but changes dramatically above 1300°C, indicating that $T_{c1} \approx 1290$°C, rather than $T_{c2} \approx 550$°C transition, is directly relevant to the vortex density. Figures 4a-4d show our superlattice DF-TEM images in a $Nd_2SrFe_2O_7$ single crystal with various cooling rates across $T_{c1}$. The full spatial maps are shown in Supplemental Material section III, Fig. S2 and Fig. S3 [31].

Indeed, the vortex density varies in a systematic manner, following the universal behavior of the Kibble-Zurek scaling mechanism for cooling rates $t$ across $T_{c1} \approx 1290°C$ in the range of 5 to $7000°C/h$. The largest vortex density of about $21.6 \, \mu m^{-2}$ is experimentally observed at the highest cooling rate value as shown in Fig. 4a. The cooling-rate dependence of the vortex density follows a power law $\{D \propto t^n\}$ with the exponent $n$ of 0.59 (Fig. 4d inset), reminiscent of the value for multiferroic hexagonal $RMnO_3$ (R= rare earths) [7].

Type-II $Z_8$ T' (anti-)vortices span the whole bulk as observed not only in the *ab* plane, but also in an *ac* surface as shown in Fig. 4e. Domain walls tend to elongate along the *c*-axis direction, probably due to the translational shift of a half *c*-lattice at DWs. An extended three-dimensional (3D) picture is illustrated in Fig. 4f. In a two-dimensional (2D) surface (right panels in Fig. 4f), four domains and four DWs consisting of two types of O'-symmetry DWs (light-yellow and light-red curves in Fig. 4f) meet at a core (brown and green circles in Fig. 4f). In three-dimensional (3D) space, those cores converge to a vortex line (black line in Fig. 4f). By symmetry, the tilt axes of the T' domains rotate by 90 degrees in adjacent bilayers (red arrows in Fig. 4f), so an antivortex is always right beneath a vortex in the adjacent bilayers, and vice versa, and vortex and antivortex alternate along a vortex line.

## IV. IN-SITU HEATING DOMAIN OBSERVATION

To investigate the possible vortex-to-vortex transformation that we discussed earlier, we further performed in-situ TEM heating experiments above $T_{c2}$ on $Nd_2SrFe_2O_7$ single crystals. Two specimens with initial low (Figs. 5b-5c) and high (Figs. 5e-5f) type-II $Z_8$ T' vortex densities, resulting from different cooling rate across $T_{c1}$, were examined. Upon heating, the $S_2 = \frac{1}{2}(200)_T = (110)_{T'}$ spots begin to fade (Fig. 5a) while the $S_1/S_3$ spots become stronger over a wide temperature range, indicating the hysteretic phase transformation from the T' to O'$_{Tc2}$ phases. In real space, four-level colors exist; rectangular domains with dark- and light-grey contrasts appear above $T_{c2}$ (Figs. 5b-5c), corresponding to two types of O'$_{Tc2}$ orthorhombic twins in addition to the leftover O'-symmetry DWs with dark and bright contrasts. The reverse contrasts between Figs. 5b-5c further confirm the O'$_{Tc2}$ twin nature. Analogously, the O'$_{Tc2}$ orthorhombic twins form at 600°C in the specimen with a higher $Z_8$ T' vortex density, in a few hundred-nm in length though (Figs. 5e-5f). The density of O'$_{Tc2}$ orthorhombic twins reveals a positive correlation with the initial type-II $Z_8$ T' vortex density, i.e. the cooling rate across $T_{c1} \approx 1290°C$.

At first sight, instead of seeing the vortex-vortex transformation (left path in Fig. 1b), lamellar twins appear above $T_{c2}$. A carefully analysis of color levels in Figs. 5b-5c reveals the formation of $Z_4$ O'$_{Tc2}$ vortices which consist of four O' states, i.e. $\varphi = \frac{n\pi}{4}$ ($n = 1, 3, 5, 7$) meeting at one point when considering the finite-width O' DWs as narrow domains. Figure 5d is the vector map derived from Fig. 5b-5c. For sake of simplicity, two colors (light-red and light-yellow) instead of four colors are applied for the four states of O'$_{Tc2}$ phase. No T' phase exist. The new nucleated O'$_{Tc2}$ phase contains a larger magnitude of the tilt component than the O' phase of original DWs. The in-situ DF-TEM analysis and structural factors with different octahedral tilting amplitudes are shown in Supplemental Material Figs. S4-S5 [31]. The same extinction condition and the weaker domain contrast of the new nucleated O'$_{Tc2}$ phase confirms our assumption.

Figure 6b illustrates two examples of $Z_4$ O'$_{Tc2}$ vortex domains observed in Fig. 5e-5f. An appealing aspect is that those T' vortex cores remain immobilized and become the $Z_4$ O'$_{Tc2}$ vortex cores (Fig. 6b) upon heating. Note that the vortex cores may stay either at the twin boundaries (upper figures of Fig. 6b) or within a twin domain (lower figures of Fig. 6b). Distinct from our proposed type-II $Z_8$ O' vortex or lamellar O' twins and (left or right paths in Fig. 1b), surprisingly, $Z_4$ O'$_{Tc2}$ (anti)vortices develop with the leftover O'-symmetry DWs captured within the nucleated O'$_{Tc2}$ twins. In addition, partial (anti)vortices defined by an uncompleted cycle of vectors (order parameters) may form at the intersection of a nucleated O'$_{Tc2}$ twin boundary and a O' wall as clearly depicted in Fig. 5d and Fig. 6c.

The transition from $Z_8$ T' vortices to $Z_4$ O'$_{Tc2}$ vortices is further explained by the evolution of energy landscapes across $T_{c2}$. At room temperature, the T' phase is stable and the O' phase is metastable as shown in Fig. 6a. To reduce the interfacial energy, the stable vortex core structure is $Z_8$ T' vortices with the O' phase within the DWs. On the other hand, when the temperature is higher than $T_{c2}$, the T' phase becomes the O'$_{Tc2}$ phase while the O' phase almost remains unchanged in the energy surfaces in Fig. 6a. Due to the large energy barriers between the O'$_{Tc2}$ and O' phase, the O' phase maintains within the DWs as shown experimentally (Fig. 5). In contrast to the T' phase, the O'$_{Tc2}$ phase displays spontaneous distortion, and thus the domain structure shows straight twin boundaries. Note that the O'$_{Tc2}$ and O' phases are isosymmetric and their difference is on the magnitude of structural distortions.

## V. SPONTANEOUS DISTORTION EFFECT ON DOMAIN PATTERNS.

Since type-II $Z_8$ T' topological defects are associated with octahedral distortions, we turn our attention to the effects of spontaneous distortion on the vortex formation. It is well known that the amplitude of local oxygen octahedral distortions in RP compounds can be manipulated by varying the ionic size mismatch (i.e. the so-called tolerance factor [38]). We have prepared high quality polycrystalline samples of the solid solution $Nd_2Ca_xSr_{1-x}Fe_2O_7$ in the range of $0 \leq x \leq 0.6$, with the aim of increasing the tilt amplitude by the partial chemical substitutions of small $Ca^{2+}$ for large $Sr^{2+}$. We establish the phase diagram (Fig. 7a) and spontaneous distortion (Fig. 7b) of $Nd_2Ca_xSr_{1-x}Fe_2O_7$, as inferred from the in-situ and ex-situ domain observations during various heat treatments (Supplemental Materials section IV, Figs. S6-S7 [31]). For example, we have observed orthorhombic twin-like structures at $x = 0.5$ ($Nd_2Ca_{0.5}$-S in Fig. 7c) at room temperature. Twins rearrange completely after the heat treatment up to 1470°C ($Nd_2Ca_{0.5}$-Q in Fig. 7d) while little changes with various thermal treatments below 1470°C. Here, capital Q and S denote quenching and slow cooling processes, respectively, across both $T_{c1}$ and $T_{c2}$. Surprisingly, the electron-diffraction analysis reveals a persistent tetragonal T' phase all the way to the composition $x = 0.5$ at room temperature (see Supplemental Material Fig. S6 [31]). On the other hand, the existence of spontaneous distortion is evident, showing the peak broadening and splitting as a function of Ca content in X-ray powder diffraction pattern (Fig. 7b). The spontaneous distortion changes with different heat treatments (bottom panel of Fig. 7a). The spontaneous distortion of $\Delta a/a = (b\text{-}a)/a$, where $a$ and $b$ are the orthorhombic lattice parameters, varies between 0 and 1.0 % with the Ca substitution. $Nd_2Ca_{0.5}$-S has the largest 1% spontaneous distortion and the finest twin domains (Fig. 7c) while $Nd_2SrFe_2O_7$ possesses zero distortion.

Figures 7e-7f reveal microstructures of $Nd_2Ca_{0.5}$-S and $Nd_2Ca_{0.5}$-Q using superlattice DF-TEM images. A drastic change occurs in $Nd_2Ca_{0.5}$-S (Fig. 7e) wherein no type-II $Z_8$ T' vortex has been observed. The matrix with the bright contrast is still the T' phase while DWs reveal dark. High-density dark wiggled loops self-group and elongate along either $[110]_T$ or $[\bar{1}10]_T$ diagonal direction. Those regions of elongated DWs alternate and form a quasi-periodic array over macroscopic distances and pseudo twins (Fig. 7e and Supplemental Material Fig. S8 [31]). The origin of these elongated closed loops becomes clear by a complementary domain-wall mapping (Supplemental Material section V and Fig. S9c-9g [31]). Our DF-TEM images with different superlattice peaks indicate a similar orthorhombic-like local structure of those loops, i.e. T'/O' loops, loop domain walls with O' distortions in the T' background/domains. In addition, the elongated direction is associated with two different tilt axes. Therefore, orthorhombic twin-like structure observed under a polarized optical microscope (Fig. 7c-7d and Supplemental Material Fig. S8 [31]) is, in fact, associated with pseudo twins in T' matrix with alternating two types of T'/O' loops: each type of orthorhombic loops conglomerate periodically. Emphasize that this is highly unusual, since the matrix is T' phase and no twinning would be expected in a tetragonal structure in general.

A similar pattern was found in $Tb_2Ca_{0.65}Sr_{0.35}Fe_2O_7$ (Fig. 7g), where those T'/O' loops tend to sharpen up and avoid mutual intersections. As shown in the inset of Fig. 7g, inside and outside a loop, the tilt directions of two T' domains always rotate $90^o$ (red arrows). Note that the background matrix dominant in terms of volume is the T' phase and one out of four T' state is favored in an extended region (red arrows outside the loops). Interestingly, the partially restoration of type-II $Z_8$ T' vortices occurs in $Nd_2Ca_{0.5}$-Q with a fast-quenching process across $T_{c1}$, resulting in a low density of pseudo twins (Fig. 7d). The size of local patches varies in a wide range from tens of nanometers to submicron. Inside the submicron domains of one type of T'/O' loops, there still contain small patches of the other type of T'/O' loops. A complete analysis can be found in Supplemental Material Figs. S10c-10e [31]. Those local patches can be considered as nano-twins and should behave invisible under a polarized optical microscope, which explains the reduction of optically observable twin-like structure in Fig. 7d.

We then discuss the underlying mechanisms behind the formation of vortices and T'/O' loops based on phase-field simulations (Fig. 8a-8b). When the temperature is between $T_{c1}$ and $T_{c2}$, the stable domain structure is orthorhombic twins with the possible coexistence of $Z_4$ O'$_{Tc2}$ vortices. For a slow cooling across $T_{c1}$ and large spontaneous distortion, the $Z_4$ O'$_{Tc2}$ vortex density is very low. For simplicity, the initial domain configuration in our simulation is an O'$_{Tc2}$ twin as shown in Fig. 8a, with the two domains correspond to $\varphi = 1\pi/4$ (light red) and $7\pi/4$ (light yellow), respectively. When the temperature is cooled below $T_{c2}$, T' domains will replace O'$_{Tc2}$ domains. According to the energy landscape shown in Fig. 1c, the yellow T' domain ($\varphi = 8\pi/4$) naturally becomes the majority matrix since the corresponding order parameter is close to the two initial O'$_{Tc2}$ domains. The red ($\varphi = 2\pi/4$) and green ($\varphi = 6\pi/4$) T' domains grow as inclusions within the matrix, and the boundaries of the inclusion are O' nano domains (Fig. 8b), consistent with the experimental observation (Fig. 7e and Supplemental Material Fig. S9 [31]). The full evolution process is demonstrated in supplemental Movie [31]. The elimination of blue T' domain ($\varphi = 4\pi/4$) can be understood as a long switching path or a high energy barrier across the center peak (Fig. 1c). Evidently, the enhanced spontaneous distortion by chemical substitutions of $Ca^{2+}$ for $Sr^{2+}$ favors well-defined large-scale lamellar O'-orthorhombic twins at the intermediate state and results in the transformation from type-II $Z_8$ T' vortex to T'/O' loops (Fig. 8c). Note that, in $Na_2Ca_{0.4}$ within a moderate spontaneous distortion, one can completely manipulate the T'/O' loops (Supplemental Material Figs. S9b-9d) or type-II $Z_8$ T' vortices (Figs. S10a-10b) state with different heat treatments.

## VI. Conclusions

In summary, we discovered and explored the formation of topological defects of type-II $Z_8$ T' (anti)vortices at room temperature, O'$_{Tc2}$ $Z_4$ (anti)vortices at high temperatures, and T'/O' loops in the presence of significant spontaneous distortion, traversing successive second-order and first-order phase transitions in a bilayered ferrite. We demonstrate that the dynamics at the second-order critical temperature plays the decisive role in determining the final density of topological type-II $Z_8$ (anti)vortices with a power-law dependence, following the Kibble-Zurek scaling mechanism. Upon heating across $T_{c2}$, the vortex cores are not affected by new symmetry breaking and become the cores of $Z_4$ O'$_{Tc2}$ (anti)vortices. The coexistence of O' phases with different distortion amplitudes during in-situ heating implies a locally preserved structure and hysteretic and complex fluctuations/variations of the $X_3^-$ mode order-parameter angles in the first-order type $T_{c2}$ transition. The expected vortex-to-vortex transformation outlined in the left path of Fig. 1b was not observed in our system, but may require a system with two continuous (i.e., second-order) phase transitions. In addition, unprecedented type-II $Z_8$ T' (anti)vortex to T'/O' loop evolution, occurring with enhanced spontaneous distortion, represents a finely balanced set of tilt order parameters, which may enable external thermodynamic and strain/pressure control. It should be further investigated how these real-space crystallographic order parameter topologies presented here influence the configurations of antiferromagnetic domains and DWs through spin-lattice coupling in this bilayer ferrite. Furthermore, there are numerous materials undergoing multiple structural and/or magnetic phase transitions, our findings is a stepping stone to study the evolution of topological defects in any of those materials.

## ACKNOWLEDGMENTS

The work at Rutgers was funded by the Gordon and Betty Moore Foundation's EPiQS Initiative through Grant GBMF4413 to the Rutgers Center for Emergent Materials. The work at Penn State was supported as part of the Computational Materials Sciences Program funded by the US Department of Energy, Office of Science, Basic Energy Sciences, under Award Number DE-SC0020145.

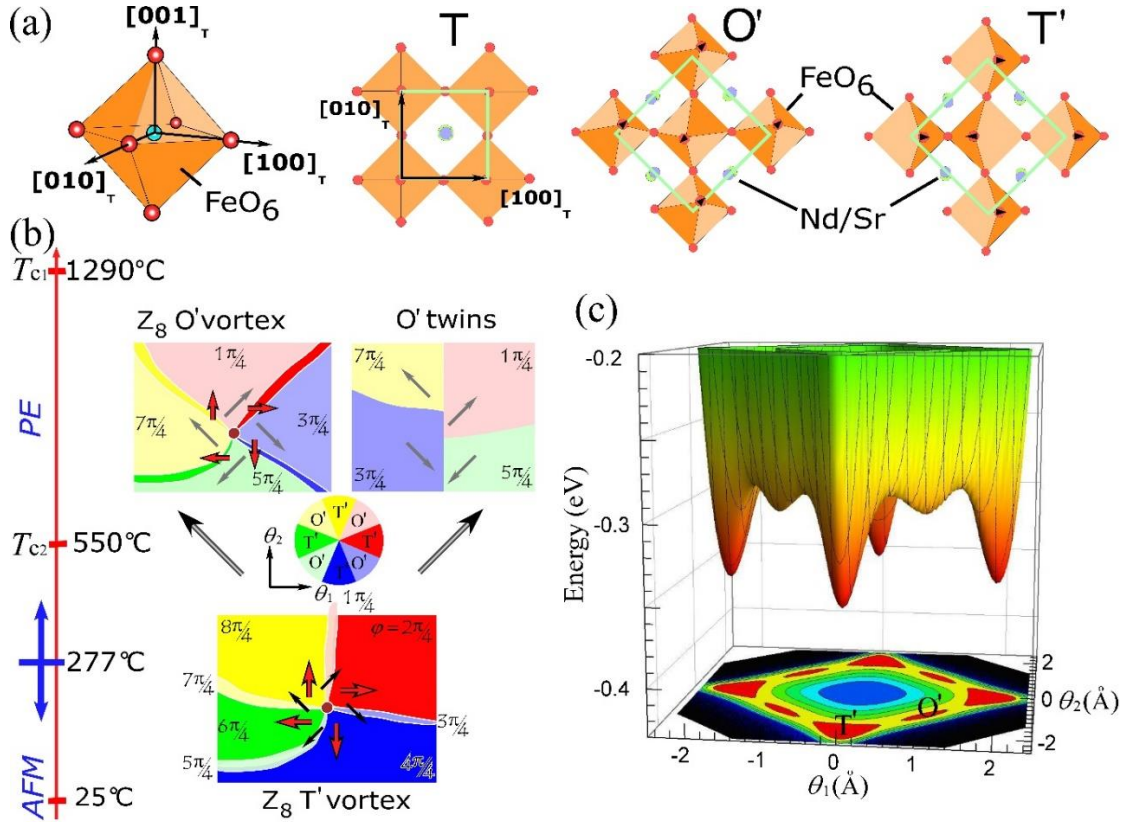

FIG. 1 (single-column color online) (a) Two sequential phase transitions of $Nd_2Ca_xSr_{1-x}Fe_2O_7$, high temperature T ($I4/mmm$)➔ O' ($Amam$)➔ room temperature T' ($P4_2/mnm$), associated with various tilts of the $FeO_6$ octahedra. Black arrowheads indicate directions of apical oxygen (red spheres) displacements. (b) Reported $T_{c1}$ (T to O') and $T_{c2}$ (O' to T') transitions in $Nd_2SrFe_2O_7$. Type-II $Z_8$ T' vortex forms at room temperature. Left path illustrates a type-II $Z_8$ T' vortex to a type-II $Z_8$ O' vortex transformation through T' domains narrowing (dark colors) and O' walls broadening (light colors) across $T_{c2}$. The transformation occurs around an immobilized brown vortex core. Right path depicts the non-vortex transformation by developing orthorhombic O' twins across $T_{c2}$. A color wheel illustrates the color assignment based on the order parameter direction and the corresponding phase $\varphi = \frac{n\pi}{4}$ ($n = 1-8$). Red and black arrows denote the T' and O' phases, respectively. Grey arrows signify the O' phase under a stimulus and it may or may not share the same tilt magnitude as the black ones. (c) Energy landscape for the doped $Nd_2SrFe_2O_7$ system, which is fitted based on the DFT calculations at 0 K. Note that the presence of four global minima of T' phase in the brim and four local minima of O' phase between them. $\theta_1$ and $\theta_2$ denote the $X_3^-$ mode tilt of oxygen octahedron.

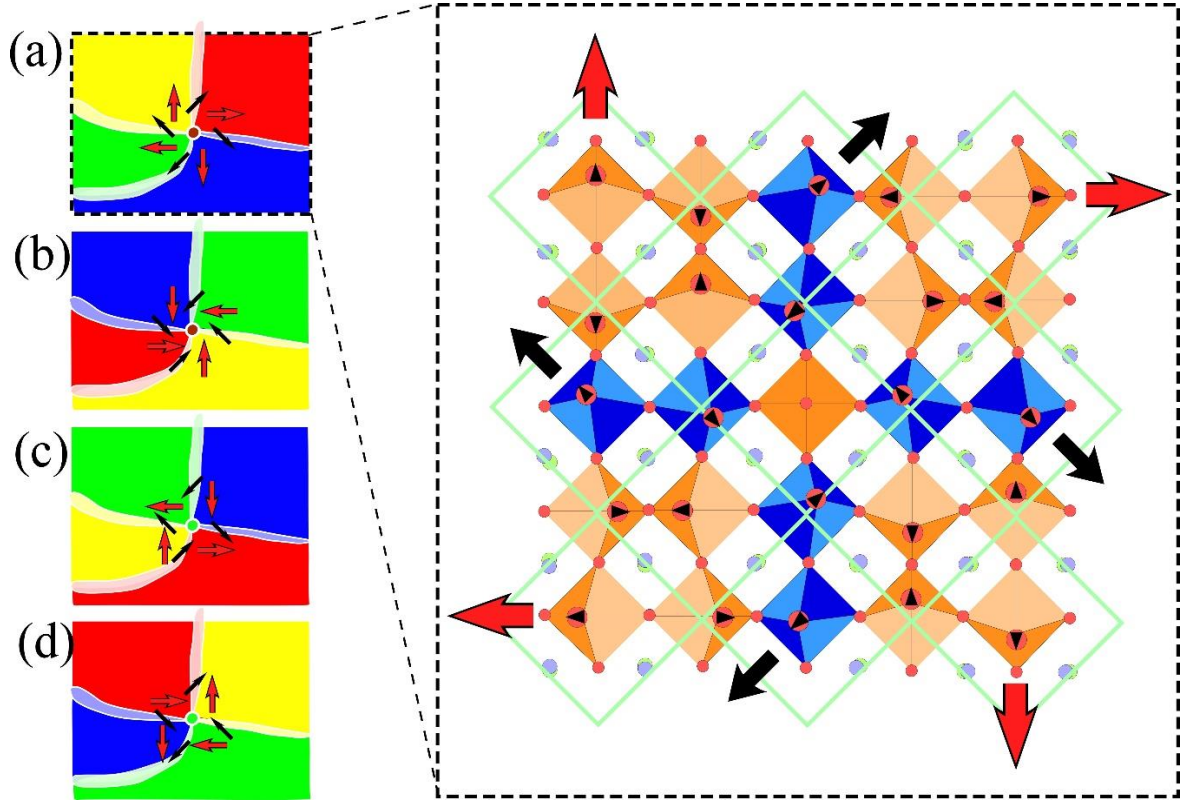

FIG. 2 (single-column color online) (a-b) Two types of vortex-like domains and (c-d) two types of antivortex-like domains observed in Fig. 3f. The schematics in the right panel illustrate the local distortion of the $Z_8$ vortex. Black arrowheads indicate directions of apical oxygen (red spheres) displacements. The green squares for the $\sqrt{2}\times\sqrt{2}$ supercells demonstrate the continuous rotation with various octahedral tilts around an un-tilted core showing the T symmetry.

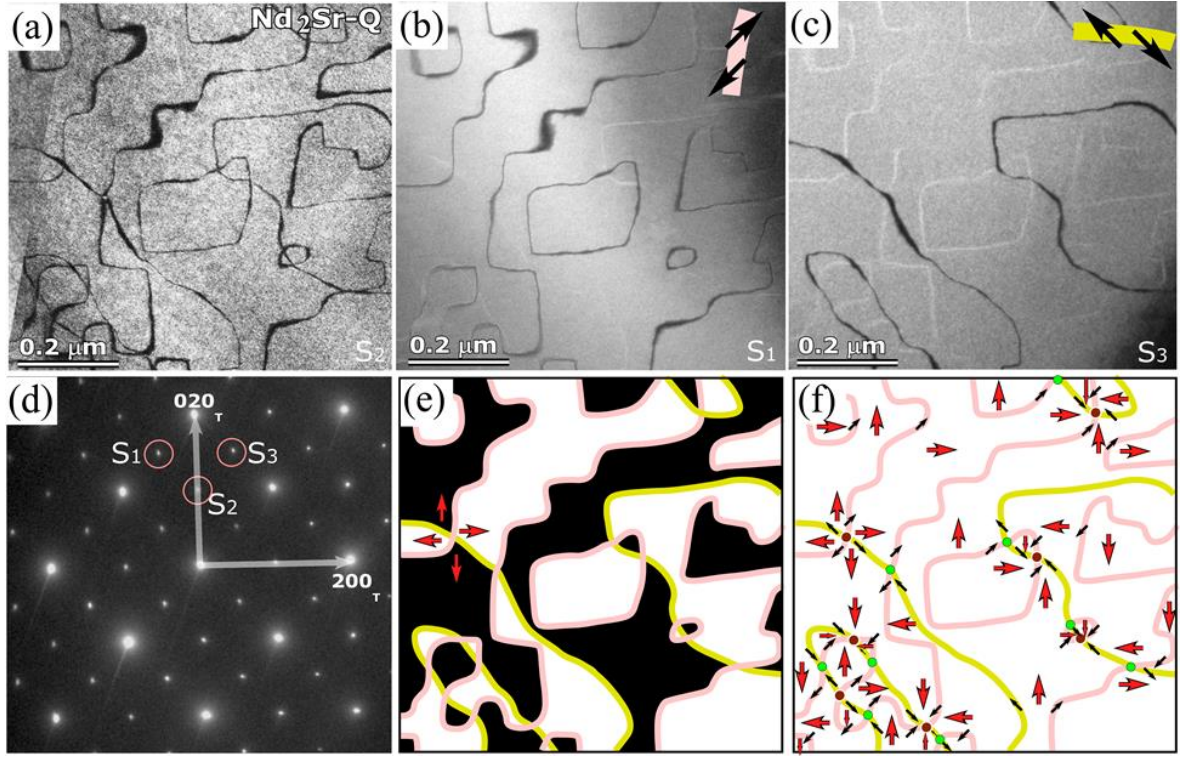

FIG. 3 (single-column color online) DF-TEM images of Nd$_2$SrFe$_2$O$_7$-Q taken under different tilting conditions, showing antiphase boundaries, which appear as dark lines with spots (a) S$_2$ = ½(200)$_T$ = (110)$_T'$, (b) S$_1$ = ½(3$\bar{1}$0)$_T$ = (120)$_T'$, and (c) S$_3$ = ½(310)$_T$ = (210)$_T'$. (b-c) reveal only a part of DWs in (a), indicating the inequivalent nature of those DWs. Octahedral tilting directions in real space are of those dark lines are shown in the up-right corner. (d) Selected area electron diffraction (SAED) pattern along [001] projection. (e) A cut-through graph can be readily constructed. Light-yellow lines never cross any light-yellow lines and light-red lines never cross any light-red lines. Specifically, it is a cut-through graph with two types of DWs. Two-proper coloring, black and white, is sufficient to identify the domains without the neighboring domains sharing the same color. Black and white T' domains are related with [100]$_T$ and [010]$_T$ tilt axes (red arrows). (f) The full mapping of eight variants in the domains and DWs and the full identification of vortices (brown) and antivortices (green). A vortex is always surrounded by antivortices connected with DWs and vice versa.

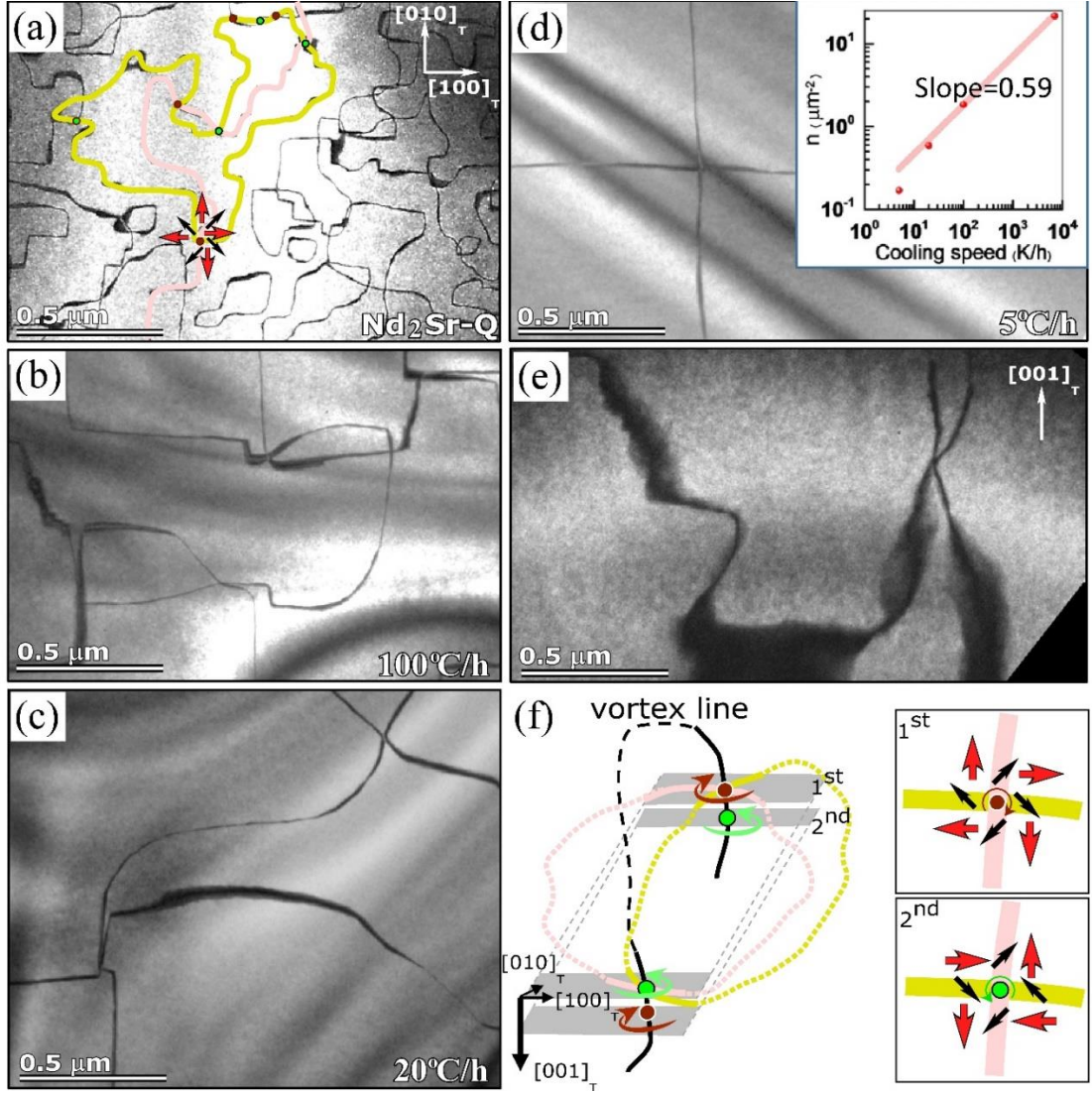

FIG. 4 (single-column color online) Cooling rate dependent type-II $Z_8$ T' vortex domain patterns in $Nd_2SrFe_2O_7$ crystals. Light-yellow and light-red bold lines represent DWs of the O'-symmetry originated from $[110]_T$ and $[\bar{1}10]_T$ tilt axes. In-plane DF-TEM images of crystals cooling from 1300ºC (above $T_{c1}$) by (a) fast quenching (7000ºC/h across $T_{c1}$), (b) with a cooling rate of 100ºC/h, (c) 20ºC/h, and (d) 5ºC/h. The inset shows the cooling rate dependence of the vortex density (vortex number/area), exhibiting the Kibble-Zurek like increase of vortex density with cooling rate. (e) Side-view DF-TEM image showing an elongated type-II $Z_8$ T' vortex. (f) The schematic of 3D vortex line connecting cores of vortices (brown circles) and anti-vortices (green circles). Black, light-red and light-yellow curved lines indicate vortex line and two types of DWs, respectively. The right panel illustrates the 90º domain relation (red arrows) between adjacent bi-layers along $c$-direction.

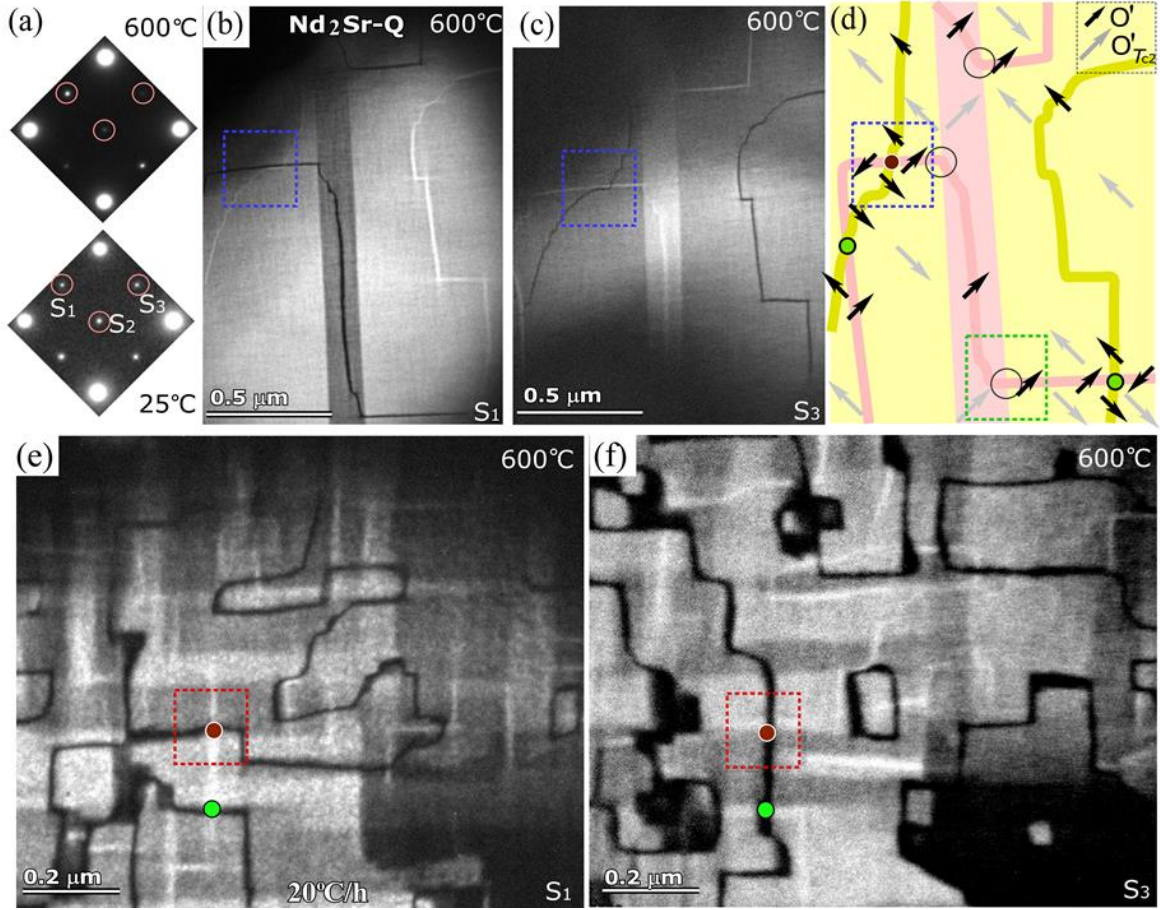

FIG. 5 (single-column color online) In-situ heating domain patterns in Nd$_2$SrFe$_2$O$_7$ crystals. (a) SAED patterns at 25°C and 600°C (above $T_{c2}$), showing the T' to O' transition. Note that S$_2$=1/2(200)$_T$-type spots are not allowed in the O' phase. The residual S$_2$ intensity reflects the 1$^{st}$ order nature of the transition. High temperature DF-TEM images of (b-c) Nd$_2$Sr-Q and (e-f) Nd$_2$Sr-20°C/h, taken using superlattice S$_1$ and S$_3$ spots. Four levels of color are identified: black, dark-grey, light-grey and white. Black and white correspond to leftover O'-symmetry DWs while dark(light)-grey contrasts relate to the new nucleated O'$_{Tc2}$ domains. The reversed contrast of those emerging rectangular dark- and light-grey O'$_{Tc2}$ domains indicates 90°-crystallographic orthorhombic twin relation. (d) The corresponding vector map of (c-d), showing Z$_4$ O'$_{Tc2}$ (anti)vortices. Light-red and light-yellow blocks are O'$_{Tc2}$ twin domains. Black and grey arrows represent the O' phase with different tilt magnitudes. Brown and green circles depict immobilized vortex and antivortex cores. Partial (anti)vortices defined by an uncompleted cycle of vectors (order parameters) are circled. The domain evolutions in red and green dotted rectangles are shown in Figs. 6b-6c.

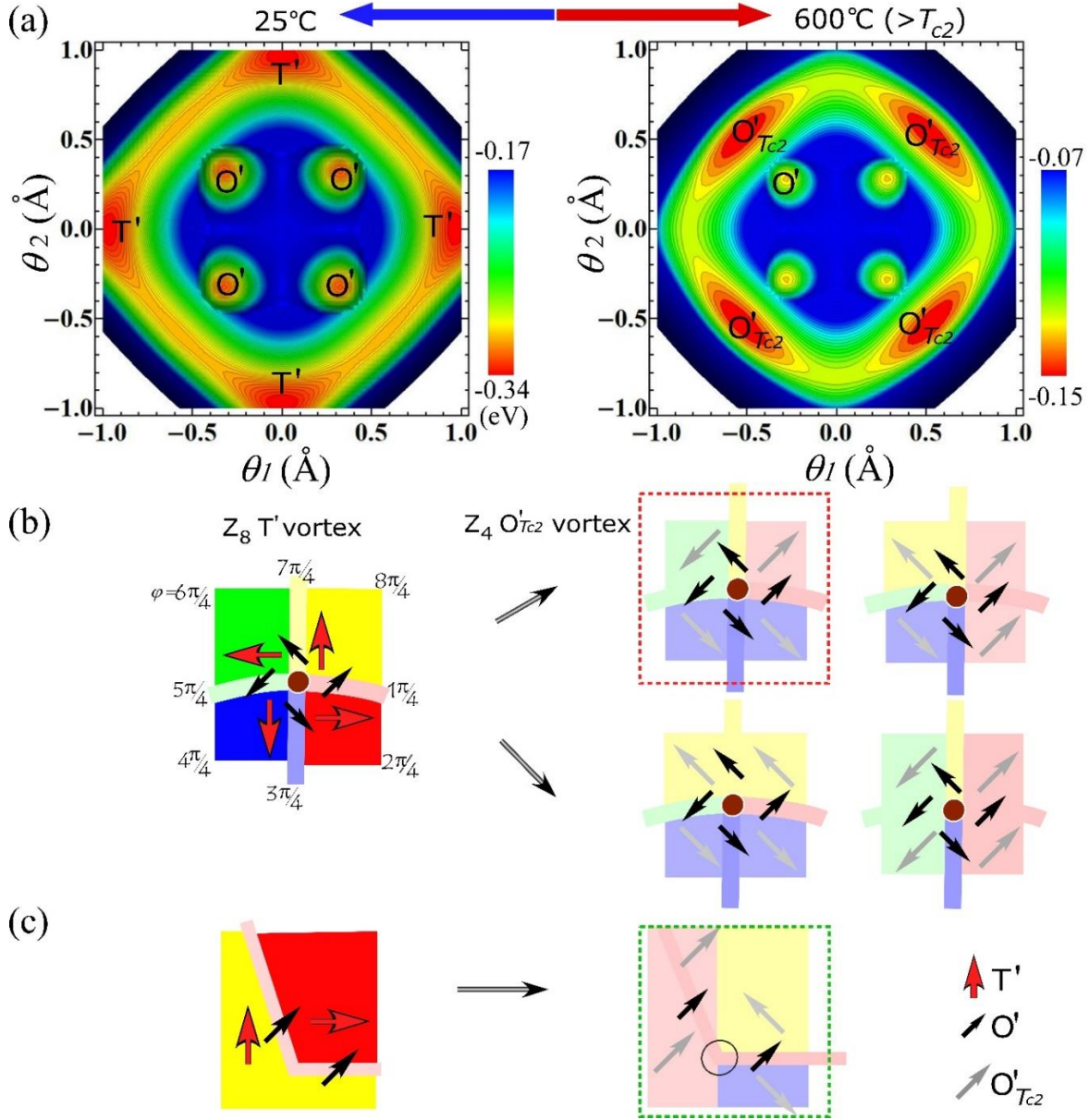

FIG. 6 (single-column color online) Schematics for the energy contours and domain structure evolutions. (a) Energy landscapes with the temperature below and above $T_{c2}$. At 25°C (below $T_{c2}$) T' corresponds to the stable state while O' is metastable. At 600°C (above $T_{c2}$), O'$_{Tc2}$ becomes stable, and O' remains metastable. The energy landscapes are schematic plots for the energy surfaces at 25°C and 600°C, which are suggested from our TEM images at those temperatures. (b) Schematic examples of $Z_4$ O'$_{Tc2}$ vortices and the evolutionary scheme. The color assignment is based on the order parameter direction as shown in Fig. 1b. Red, black and grey arrows represent T', O', and O'$_{Tc2}$ phases. (c) A partial vortex arises at the intersection of O'$_{Tc2}$ twin boundary and one leftover O'-symmetry DWs (nano domains).

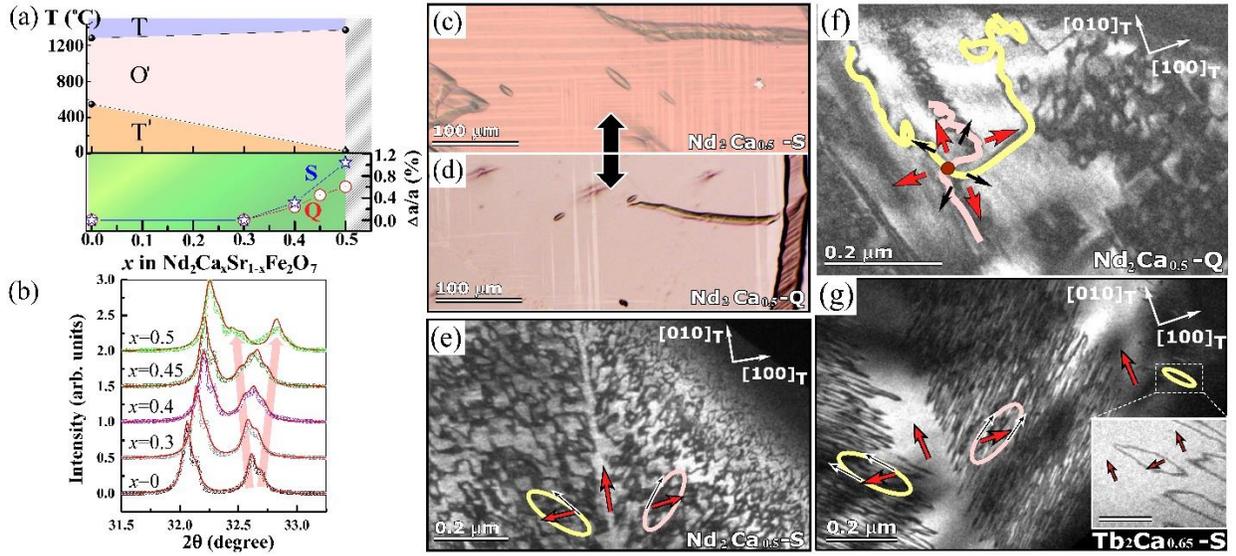

FIG. 7 (two-column color online) (a) The phase diagram and the evolution of spontaneous distortion ($\Delta a/a$) as a function of the Ca content in $Nd_2Ca_xSr_{1-x}Fe_2O_7$ with Q (red line) and S (blue line) heat treatments. The capital Q and S signify quenching and slow cooling with the rate of 100ºC/h from 1400-1470ºC (above $T_{c1}$). Spontaneous distortion ($\Delta a/a$) changes with different heat treatments which is close related with the fine domain structure shown in (c-d). (b) $X$-ray data showing the peak broadening and splitting of the $(100)_T/(010)_T$ reflection as a function of Ca content. Raw and refined data are shown in color circles and red lines, respectively. Reflection-mode polarized optical microscope images of (c) $Nd_2Ca_{0.5}$-S with pseudo twins about tens of μm in size and (d) $Nd_2Ca_{0.5}$-Q with a low density of pseudo twins at room temperature. In-plane DF-TEM images of (e) $Nd_2Ca_{0.5}$-S and (f) $Nd_2Ca_{0.5}$-Q, showing distinct microstructures at room temperature. The bright and dark contrasts are associated with T' domain and O' DWs, respectively. $Nd_2Ca_{0.5}$-S of 1% spontaneous distortion reveals periodic O' DWs while $Nd_2Ca_{0.5}$-Q of 0.6% spontaneous distortion shows local aggregation of high-density O' DWs. (g) The microscopic image of $Tb_2Ca_{0.65}Sr_{0.35}Fe_2O_7$ of 0.7% spontaneous distortion, showing similar pseudo twins as $Nd_2Ca_{0.5}$-S. Inset shows a blow-up image of T'/O' loops with 90º rotation of the neighboring T' domains. Scale bar: 0.2 μm.

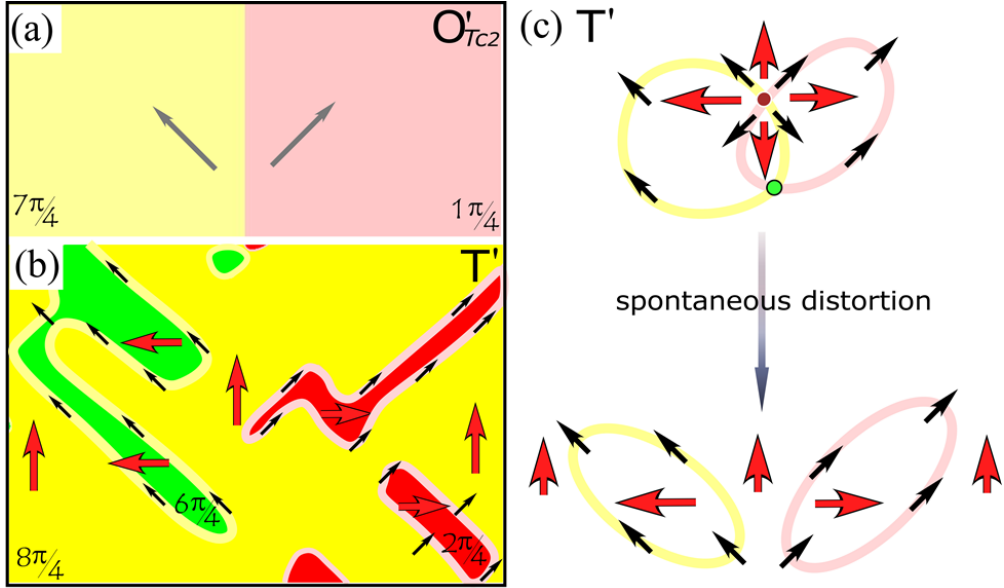

FIG. 8 (single-column color online) (a-b) Phase-filed simulation for the domain structure evolution across the $T_{c2}$ transition, starting from (a) initial high-temperature O'$_{Tc2}$ twin domains to (b) low-temperature T'/O' loops with $\varphi = 8\pi/4$ (yellow) as the major T' phase. Order parameter direction is denoted by arrows and colors as shown in Fig. 1b. (c) Schematics of the formation of type II $Z_8$ T' vortices and T'/O' loops in the presence of spontaneous distortion at room temperature.